# Scaling Laws in Cancer: The *Dynamic* Behaviour of the Power Exponent *p*

Caterina Guiot [1,2], Pier Paolo Delsanto [2,3], Yuri Mansury [4] and Thomas S. Deisboeck [4,*]

**Affiliations:** [1] Dip. Neuroscience, Università di Torino, Italy and [2] INFM, sezioni di Torino Università e Politecnico, Italy ; [3] Dip. Fisica, Politecnico di Torino, Italy; [4] Complex Biosystems Modeling Laboratory, Harvard-MIT (HST) Athinoula A. Martinos Center for Biomedical Imaging, Charlestown, MA 02129, USA.

**\* Corresponding Author:**

Thomas S. Deisboeck, M.D.
Complex Biosystems Modeling Laboratory (CBML)
Harvard-MIT (HST) Athinoula A. Martinos Center for Biomedical Imaging
Massachusetts General Hospital-East, 2301
Bldg. 149, 13th Street
Charlestown, MA 02129
Tel:    617-724-1845
Fax:    617-726-5079
Email: deisboec@helix.mgh.harvard.edu

**Abstract**

We have previously reported that a 'universal' growth law with *fixed* scaling exponent *p*, as proposed by West and collaborators for all living organisms, appears to be able to describe *also* the growth of tumors *in vivo*. Here, we investigate in more detail the *dynamic* behaviour of *p* using data from the literature. We show that *p* initially *decreases* before it again *increases* up to or even beyond 3/4. These results support the notion that *p* can vary over time and yet its dynamics are *independent* from the cancer type. We argue that this behaviour reflects the underlying evolving tumorigenesis in that the *minimum* of *p* signals the emergence of a fractal-like distribution network corresponding to the 'angiogenetic switch' towards a *perfusion*-dominated nutrient supply mechanism.

**Introduction**

In a previous paper we reported that West et al.'s model of 'universal growth' (West *et al*, 2001) can also describe the growth dynamics of experimental *in vivo* tumors and, furthermore, that it can also be applied to selected data from experimental tumors as well as clinical cancer data (Guiot *et al*, 2003). In that study, the scaling exponent *p* is assumed to be *fixed* over time. Here, we introduce a modification to our tumor growth model allowing the exponent *p* to vary *dynamically and continuously* over time to reflect changes in the nutrient-supply mechanisms during tumorigenesis. Using our dynamic framework, we investigate biomedical data reported in the literature to test the validity of the proposed mechanisms.

According to West *et al* (2001), the actual mass *m* and its rate of growth, *dm/dt*, are non-linearly related:

$$dm/dt = a\, m^p \left(1 - \left(\frac{m}{M}\right)^{1-p}\right), \qquad (1)$$

where M is the asymptotic value of *m(t)* and *a* is a parameter related to the metabolic rate of the particular tumor cell line. It can be shown using Eq. (1) that *m(t)* exhibits an inflection point at *t = t'* corresponding to *m = m'*, which depends on the value of *p*. The simplest way to determine *m'* is to plot the experimental values of *m* and *dm/dt* in a cartesian plane with *m* as abscissa and *dm/dt* as ordinate. In fact the curve *dm/dt* vs *m* reaches a maximum at *m = m'*, which can be accurately identified. By assuming a dynamically-changing scaling exponent *p = p(t)*, but with a slow rate of change in the time interval around *t'*, it follows that:

$$\frac{d\mu}{dt} = b\mu^p \left[ \frac{1}{p} - \mu^{1-p} \right], \qquad (2)$$

where $\mu = \frac{m}{m'}$, $b = a\, M^{p-1}$ and

$$m' \cong p^{\frac{1}{1-p}} M. \qquad (3)$$

West *et al* argue that the main mechanism for nutrient supply is related to the fractal-like distribution network, and propose the exponent $p = 3/4$ as a universal scaling factor. For tumors, this assumption implies the presence of an angiogenetic network, hence primarily corresponds to *active perfusion* for tumor growth *in vivo*. Determining the proper value of the scaling exponent $p$, however, remains a controversial issue. For example, a recent paper (Dodds *et al*, 2001) shows that $p = 3/4$ does not yield a significantly better fit of all available data than $p = 2/3$ (which derives from a simple dimensional analysis).

## Materials and Methods

To investigate the behaviour of the power exponent $p$ in experimental *in vivo* tumors, we examine data reported in Steel (1977). In order to estimate the optimal value of $p$, instead of assuming it fixed 'ab initio', we solve Eq. (2) for $p$ using the aforementioned data to estimate values of $\mu$ and $d\mu/dt$ from the growth curves and then iteratively compute the value of $p$ for which the difference between the predicted and observed values is smallest. In principle the value of $b$, which is a biological parameter and is expected to depend on the cell line, should be independently measured.

Since direct measures are not available at the moment, we first assume a tentative constant value for *b*, and then perform a recursive, autoconsistent procedure to estimate *p* at any time. Then the value of *b* is changed and the procedure repeated until a satisfactory "best fit" is reached for both *p* and *b*.

**Results**

Using the procedure outlined above, we obtain the *p(t)* plots presented in **Fig. 1**. In this study here, we investigate three out of five cell lines of tumors growing in mice as reported in Steel (1977)[1]. Note that, *regardless* of the cancer type, the scaling exponent *p* now *changes dynamically*, i.e., it initially *decreases* to a minimum before it eventually *rises* again. From our formalism or, directly from **Fig. 1**, it appears to be possible to predict the time of onset for a *perfusion-dominated* nutrient mechanism (i.e., angiogenesis). The procedure can be further refined by reformulating the problem in terms of dimensionless rescaled variables, such as $\tau$ and *r* in Eqs. (1) and (2) of Guiot *et al* (2003), respectively, but with *p* assumed to vary dynamically as a function of time. In accordance with the results of Guiot *et al* (2003), this should strongly enhance the similarity of the plots of **Fig. 1**. Also the inflection times, which vary in the figure from 5.3 to 14.2 days, should fall once the rescaled time $\tau$ is confined within a much narrower range (from about 0.21 to 0.39, in a preliminary evaluation). The $\tau$ range for inflection may be further reduced once more uniform starting times are considered (work in progress).

**Figure 1.**

---

[1] Data from the remaining two cell lines cannot be used for this procedure since the short duration of these two experimental series does not allow properly estimating the tumor mass at the inflection point.

**Discussion**

In this study, we report several interesting findings from a tumor biology perspective. We found first that there was an initial decrease in the scaling exponent $p$, likely caused by a rapid onset of volumetric tumor growth ($V_{total}$) after *in vivo* implantation. Secondly, we argue that the development of an effective vascular supply network, in agreement with West et al.'s conjecture, does lead to the observed *increase* of $p$. We thus further hypothesize that the curve's inflection point signals the *switch* in the *dominant* nutrient-replenishment mechanism from passive diffusion to active perfusion conferred by angiogenesis. It therefore fits, that this transition occurs in the three cell lines investigated here at an average tumor diameter of 6.6 mm (± 1.6 STD) which is beyond the threshold of 2-3 mm in tumor diameter that Folkman (1971) had argued would prompt the *onset* of angiogenesis. It is thus less surprising that when fitting the data from Torres Filho et al. (1995), who implanted lewis lung carcinoma cell spheroids into the dorsal skinfold chamber of CB6 mice, we found that the inflection of $p$ occurs at approximately Day 6 post-tumor implantation, i.e., when the *vessel density* had reportedly reached already 81 percent of its maximum value. Moreover, in this study here, values for $p$ *beyond* the anticipated 0.75 may suggest that active perfusion ($p = 3/4$) is complemented by another supply mechanisms, likely by passive diffusion, when vascular density approaches its plateau phase. We further argue that such an *increased* surface diffusion can be explained by the onset of central apoptosis and necrosis, which should reduce the actively metabolizing tumor (cell) volume ($V_a$) while restricting it to the highly proliferating tumor surface, and hence effectively raises the [S/$V_a$] ratio. In addition, we found, that for the data analyzed here, this *dynamic p*-behaviour appears to be *independent* of cancer type. It is also noteworthy that the *absolute* time $\tau$ at which $p$ reaches its minimum turns out to be very *similar* across the different tumor cell lines. Together, these findings further support our conjecture about the *universality* of the tumor growth model.

Based on the presented results we argue that the scaling exponent *p* shows distinct dynamic patterns *in vivo* and that such monitoring of *p* may be of interest if the angiogenetic switch is to be exploited for diagnostic or therapeutic purposes. More specific experiments are called for in order to test these hypotheses, given their important implications for cancer research.

**Acknowledgements**

This work was supported in part by the National Institutes of Health (CA 085139) and by the Harvard-MIT (HST) Athinoula A. Martinos Center for Biomedical Imaging and the Department of Radiology at Massachusetts General Hospital. Y.M. is the recipient of a NCI-Training Grant Fellowship from the National Institutes of Health (CA09502). We also wish to thank Drs. M. Griffa and P.G. Degiorgis for useful discussions.

**References**

Dodds PS, Rothman DH, Weitz JS. (2001) Re-examination of the "3/4-law" of Metabolism. *J Theor Biol* **209**: 9-27.

Folkman J. (1971) Tumor angiogenesis: therapeutic implications. *N Engl J Med* **285**: 1182-1186.

Guiot C, Degiorgis PG, Delsanto PP, Gabriele P, Deisboeck TS. (2003) Does tumor growth follow a "universal law"? *J Theor Biol* **225**: 147-283.

Steel GG. (1977) Growth kinetics of tumours. Clarendon Press: Oxford.

Torres Filho IP, Hartley-ASP B, Borgström P. (1995) Quantitative angiogenesis in a syngeneic tumor spheroid model. *Microvascular Res* **49**: 212-226.

West GB, Brown JH, Enquist BJ. (2001) A general model for ontogenetic growth. *Nature* **413**: 628-631.

**Figure Caption**

**Figure 1.** Estimation of the scaling exponent *p* vs. time. Data from Steel (1977) refer to three different tumor cell lines implanted in mice.

**Figure 1.**

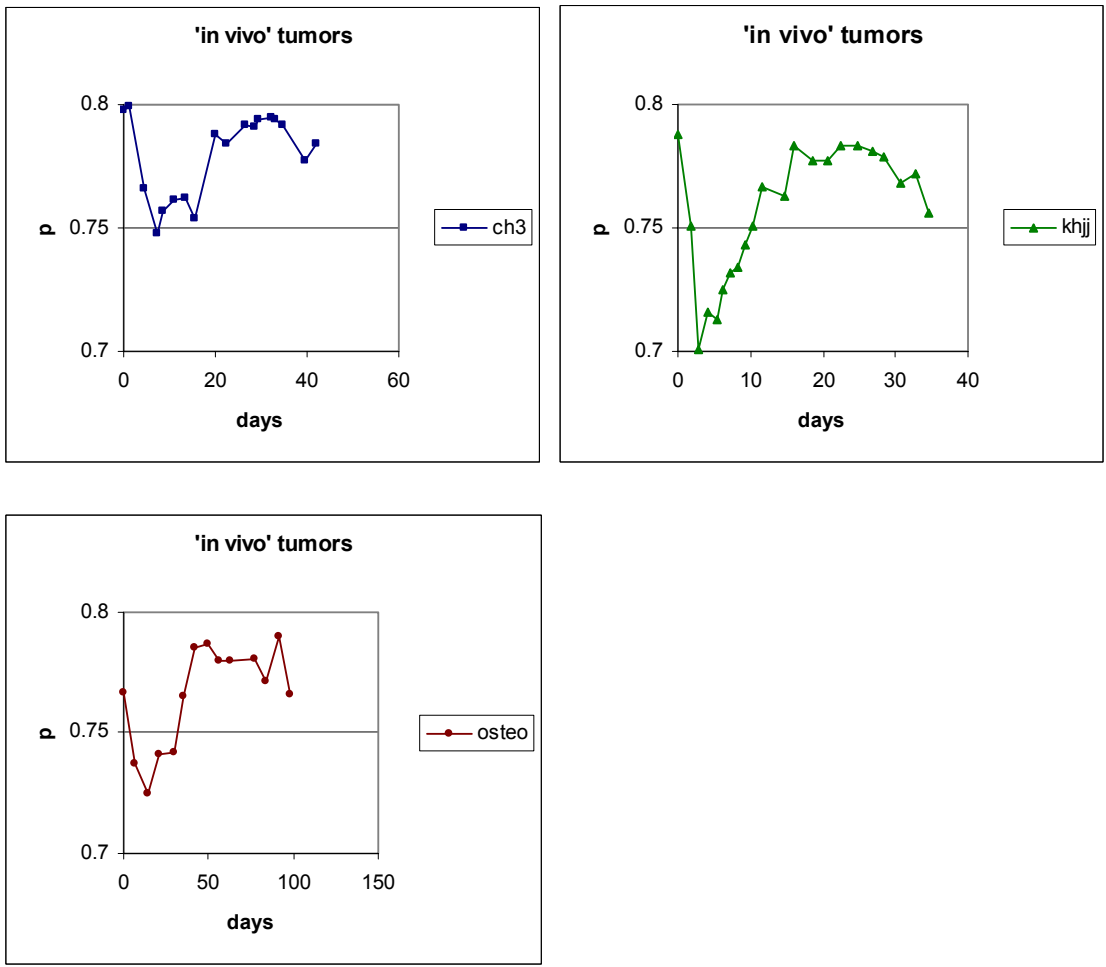